\documentclass{pasj00}
\draft

\begin{document}
\SetRunningHead{Author(s) in page-head}{Running Head}
\Received{2014/08/16}
\Accepted{2014/12/09}

\title{Near-Infrared Image of the Debris Disk around HD 15115}

\author{%
 Shoko \textsc{Sai}\altaffilmark{1,2},
 Yoichi \textsc{Itoh}\altaffilmark{2,1},
 Misato \textsc{Fukagawa}\altaffilmark{3},
 Hiroshi \textsc{Shibai}\altaffilmark{3}, and
 Takahiro \textsc{Sumi}\altaffilmark{3}
%
%
}
\altaffiltext{1}{Graduate School of Material Science, University of Hyogo,
    3-2-1 Kouto, Kamigori-cho, Ako-gun, Hyogo, 678-1297, Japan}
\email{sai@nhao.jp}

\altaffiltext{2}{Nishi-Harima Astronomical Observatory, Center for Astronomy, University of Hyogo, 407-2, Nishigaichi, Sayo-cho, Sayo, Hyogo, 679-5313, Japan}

\altaffiltext{3}{Graduate School of Science, Osaka University, 1-1 Machikaneyama, Toyonaka, Osaka 560-0043, Japan}



%


\maketitle

\begin{abstract}
We present a Subaru/IRCS H-band image of the edge-on debris disk around the F2V star HD 15115. We detected the debris disk, which has a bow shape and an asymmetric surface brightness, at a projected separation of 1--3" ($\sim$50--150 AU). The disk surface brightness is $\sim$0.5--1.5 mag brighter on the western side than on the eastern side. We use an inclined annulus disk model to probe the disk geometry. The model fitting suggests that the disk has an inner hole with a radius of 86 AU and an eccentricity of 0.06. The disk model also indicates that the amount of dust on the western side is 2.2 times larger than that on the eastern side. A several Jupiter-mass planet may exist at $\gtrsim$45 AU  and capture grains at the Lagrangian points to open the eccentric gap. This scenario can explain both the eccentric gap and the difference in the amount of dust. In case of the stellar age of several 100 Myr, a dramatic planetesimal collision possibly causes the dust to increase in the western side. Interstellar medium interaction is also considered as a possible explanation of the asymmetric surface brightness, however, it hardly affect large grains in the vicinity of the inner hole. 
\end{abstract}

\section{Introduction}


Direct imaging observations of the debris disk at visible or infrared provide the disk morphologies, which reveal the planetary formation. About 40 debris disks have been spatially resolved, and many of them have various characteristic structures such as a warp and an offset. Those features are thought to be related to gravitational interaction between the disk and planets. In fact, most of the stars with confirmed directly-imaged planets, e.g., $\beta$ Pic b (\cite{lag09}), HR 8799 bcde (\cite{mar08}, \cite{mar10}), HD 95086 b (\cite{ram13}), and HD 106906 b (\cite{bai14}) also harbor debris disks. Numerical simulations of a debris disk with a planet also demonstrated such specific features (\cite{wya99}, \cite{chi09}, \cite{kuc03}, \cite{ert12}, \cite{the12}, \cite{nes13}, \cite{rod14a}, \cite{pea14}). The detailed disk morphology from observations reveals the behavior of unseen planets. This is a powerful means of understanding planetary formation. 


HD 15115 is known as a young F2V star at a distance of 45 pc (\cite{van07}). The age of the star has a large uncertainty: if it belongs to the $\beta$ Pic moving group, the age is 12 Myr (\cite{moo06}), while the lithium abundance and X-ray emission suggest an age of several hundreds of Myr (\cite{zuc04}; \cite{rhe07}). This star has a far-infrared excess with a peak of 0.5 mJy at $\sim$70 $\micron$, which suggests the existence of a cold dust ring with a temperature of 57 K at 42 AU (\cite{zuc04}; \cite{moo11}). The dust mass is estimated to be 0.047 $M_{\earth}$ from submillimeter photometry (\cite{wil06}). 

Direct imaging observations revealed an almost edge-on debris disk stretching east to west (\cite{kal07}, \cite{deb08}, \cite{rod12}). \citet{kal07} first found the asymmetric disk structure with the Hubble Sapce Telescope (HST) and the Keck II in the V and H bands, respectively. The V-band image indicated that the east side of the disk extended to $\sim$ 315 AU, whereas the west side extended to $>$550 AU. They proposed that a nearby star, HIP 12545, had a close encounter with HD 15115 and captured disk material. \citet{deb08} presented a warped disk image with HST at the J band. They claimed that the asymmetry of the disk surface brightness (SB) was a function of wavelength, possibly owing to compositional or grain population differences between the two sides of the disk. \citet{rod12} reported Ks and L' images of a disk with a bow-like shape using Large Binocular Telescope (LBT). They proposed an explanation for the bow-shaped structure. The observed disk structure could be reproduced when the disk is inclined at 87$^\circ$ and consists of grains with a Henyey-Greenstein parameter, $g$, of 0.5. They also noted the existence of an inner hole with a radius of 1.1" (50 AU) in the disk. Moreover, the redder Ks - L' disk color on the eastern side suggested fewer 1--3 $\micron$-grains than on the western side. They attributed this disk color difference to interaction with the local interstellar medium (ISM). Recently \citet{maz14} presented H and Ks images with Gemini/NICI. Their H band image of the disk shows the existence of a symmetric inner gap at 2". The previous studies suggested hypotheses about the bow-shaped and asymmetric disk structure. To reveal the causes of these characteristic features, we directly imaged HD 15115 with high spatial resolution and clarified the disk structure by model fitting.


\section{Observations}

We obtained H-band (1.63 $\micron$) images of HD 15115 with the Subaru Telescope using the Infrared Camera and Spectrograph (IRCS) on the Nasmyth focus stage and the AO 188 adaptive optics system on UT November 11 2011. The detector was the ALADDIN I\hspace{-.1em}I\hspace{-.1em}I 1024~1024 InSb array. The pixel scale was 20.57 mas/pixel; thus, the field of view was 21.06"$\times$21.06". The sky was covered with cirrus through the observations. The H-band magnitude of the star is 5.86. The size of the stellar point spread function (PSF) was $\sim$0.1" in full width at half-maximum (FWHM), which was thought to be variable with a range of $\lesssim$0.05". 
This spatially resolution is a permissible range for detecting the disk though it is not diffraction-limited. 
We used a 0.8"-diameter coronagraph mask to reduce the flux of the central star. We observed in the angular differential imaging (ADI; \cite{mar06}) mode. The ADI mode keeps the instrument and the telescope optics aligned, while the field of view is allowed to rotate by diurnal motion. This method reduces the speckle noise and improves the contrast near the star. We obtained 82 frames; each frame was coadded with 2.0-s exposure images in four non-destructive readouts (NDRs). The number of COADDs for the first 10 frames and the last 72 frames is 20 and 10, respectively. The total integration time was 30.7 min; the rotation angle of the field of view was $\sim$17.3$^\circ$.



For photometry, we observed FS 122 with AO after observing HD 15115. The photometric standard star was taken with five-point dithering through the H-band filter. We did not use any coronagraph masks, and the stellar image was unsaturated with a PSF size of $\sim$0.1". The image of each point was coadded with two images of 2.0 s exposure in four NDRs; the total integration time was 20 s. We measured the total flux of the star with an apperture radius of 90 pixels (=1.85") for photometric calibration.


After the observations, we took dark and flat frames for basic calibrations. The flat frames were taken with a halogen lamp in the instrument. 

\section{Data reduction}

\subsection{Basic Calibration} 

 We made the following basic calibrations. First, we divided all the raw images by the numbers of COADDs and NDRs to obtain the same exposure time for all the images. Next, we average-combined eight dark images to make the dark frame and subtracted it from the science frames. Then we applied the flat-field correction. The flat frame was obtained by subtracting the averaged-combined lamp-off frame from the averaged-combined lamp-on frame and normalizing by the median of the pixel values. Finally, we removed dead/hot pixels and cosmic rays. We did not subtract sky background from the science frames. The sky was subtracted when we execute disk detection processing as discussed in Section \ref{sec3.2}.

 We determined the PSF center of each image with subpixel accuracy by seeking the origin of the point symmetry of the PSF. We made an image that was rotated by 180$^\circ$ at a certain coordinate and subtracted it from the original image. To evaluate the cleanness of the PSF subtraction quantitatively, we calculated the sum of the squared residuals in the ring area of r=50--100 pixels from the point of rotation. Each PSF center was identified when the value in the area reached a minimum. 


\subsection{Disk Detection Processing} 
\label{sec3.2}

 To differentiate a faint disk from the predominantly bright halo of the star and artifacts from the optical system such as spider patterns and speckle halo, we should remove them from each science frame. The ADI mode has artificial patterns fixed at the same position, while a disk is made to rotate. Therefore, we can make a PSF image with  artificial patterns without the disk --the PSF reference image-- from the science frames. 


We performed Karhunen-Lo\`eve Image Projection algorithm (KLIP; \cite{sou12}), which built each PSF reference image by principal component analysis (PCA). We use the pipeline\footnote[1]{http://www.sc.eso.org/~dmawet/pca-pipeline.html} which is published by Dmitri Mawet. First, we masked the inner region (r$\leqq$40 pixels = 0.82"), then divided each science image into concentric ring subsections with one pixel width. We calculated the Karhunen-Lo\`eve (KL) basis in each ring subsection. In this calculation, 
there are two tunable parameters, $n$ and $N_\delta$. $n$ is the number of PCA modes to use for building the PSF reference image. Meanwhile, $N_\delta$ is the parameter of the distance in units of PSF FWHM. The frames separated by more than $N_\delta$ each other can be used for building the PSF reference image. To build $n$ PCA modes, at least $n$ frames are necessary. In contrast, the number of the available frames decreases with increasing  $N_\delta$. Therefore, when we set a given $n$, $N_\delta$ has an upper limit to construct the PSF reference image of the $n$ frames. For example, in case of $n=5$ with the condition of this observation, $N_\delta \leqq1.054$ ($\sim$0.1") allows to use no less than five reference frames for all science images at $r>40$ pixels. We tried KLIP processing from $n=1$ to $n=40$. To keep the disk signal against self-subtraction, $N_\delta$ was set to the maximum.   We finally found that $n=15$ and $N_\delta=0.751$ maximized the disk S/N at $\sim1.5"$. In this paper, we discuss the disk of HD 15115 using the final KLIP image which was derived with these two parameters. 





\subsection{Self-subtraction Correction}
There is a possibility of self-subtraction of the disk due to the poor rotation angle of the field of view when we subtracted the PSF reference image. Several authors have explored how ADI affects recovered disk structure (e.g. \cite{tha11}, \cite{rod12}, \cite{rod14b}, \cite{mil12}). In case of LOCI, there is a technique for forward-modeling self-subtraction of spatially extended objects (\cite{esp14}). We executed self-subtraction simulations with artificial disks using those previous researches as references. We added an artificial disk on each science frame, ran the disk detection processing, and measured the brightnesses on the disk midplane and disk FWHM at each projected distance from the PSF center. We defined the brightness efficiency and the FWHM variance as the ratio of the processed to the original disk intensity and FWHM, respectively.  To have a non-biased view for the real disk, we used not a disk-like but a rectangular object for the artificial disk. We prepared nine types of rectangular disks with three intensities (0.5, 1.0, and 2.0 ADU = 0.11, 0.22, and 0.44  ${\rm mJy/arcsec^2}$) and three widths (5, 10, and 15 pix). We assume that the real disk can be expected to range within these intensities and widths. The artificial disks extend toward the same direction as the observed disk until the edge of the image. Each artificial disk was convolved with the Gaussian function of FWHM = 0.1". The self-subtraction simulations with these nine disks suggested that the both the brightness efficiency and the FWHM variance depends mainly on the disk width. Therefore, we averaged the radial profiles of brightness efficiencies and FWHM variances among the three intensities, respectively. Finally we derived the minimum and maximum at each radius among three types of averaged radial profile, yielding the scaling factors and the uncertainties of disk SB and FWHM.



\section{Results}

\begin{figure}
	\begin{center}
	\FigureFile(85mm,85mm){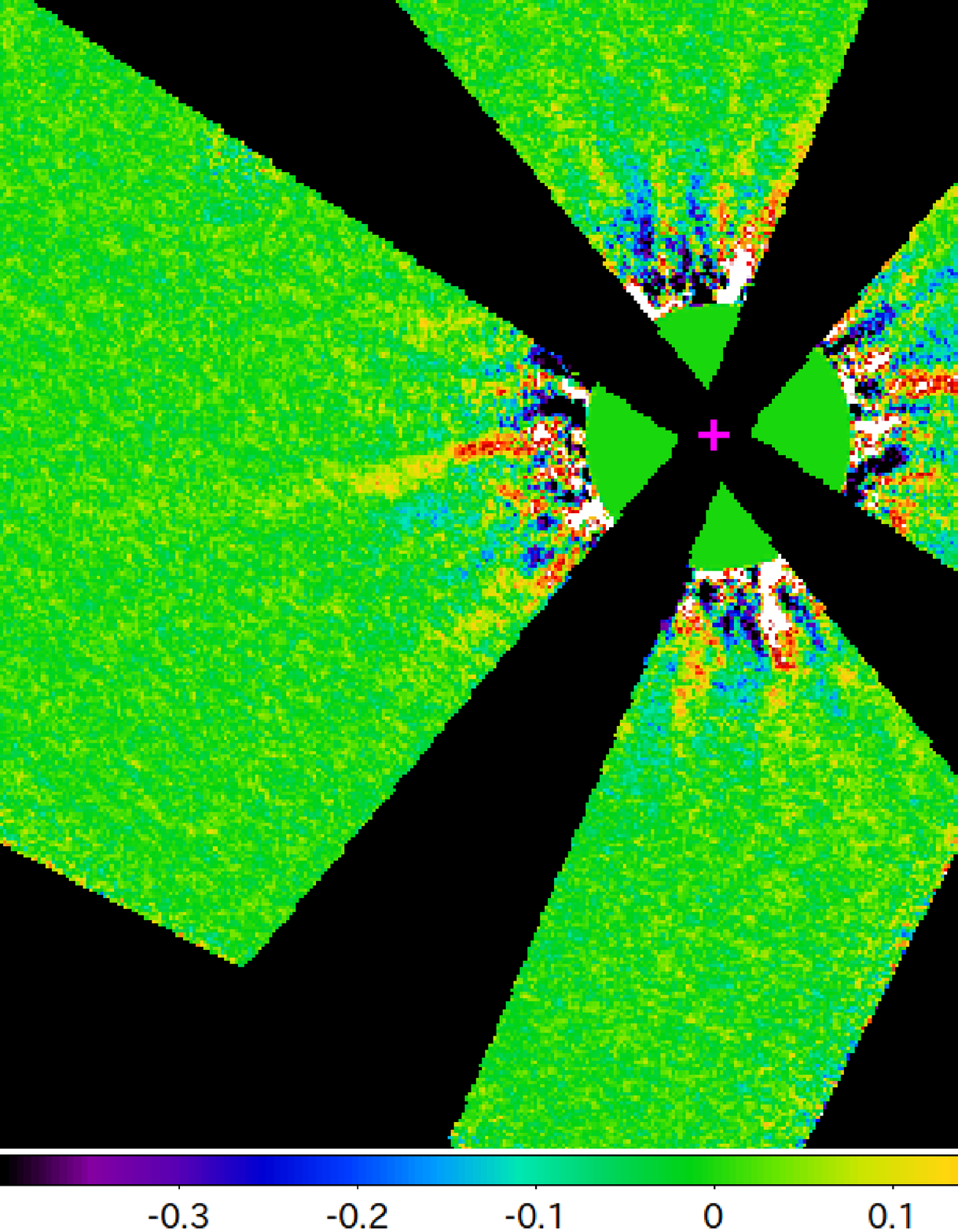}
	\caption{Subaru/IRCS H-band image of HD 15115. The image is 8.4" on a side and oriented with north up and east to the left. The black crossing patterns represent regions of the spiders passing. The edge-on disk of HD 15115 stretches east to west. A plus sign represents the position of the central star. The image is in units of ${\rm mJy/arcsec^2}$. \label{Fig1}}
	\end{center}

	\begin{center}
	\FigureFile(85mm,85mm){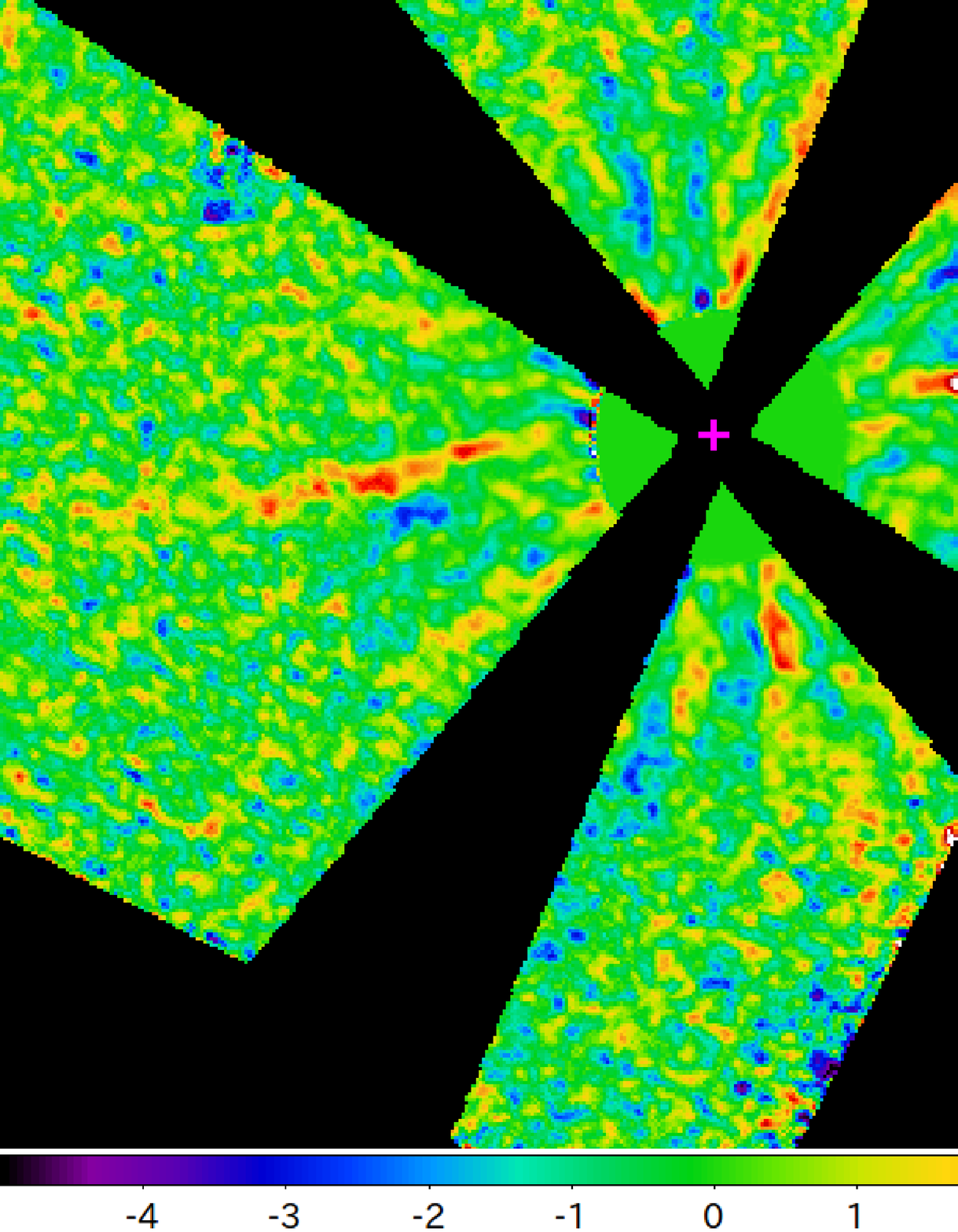}
	\caption{S/N map of HD 15115. The image is 8.5" on a side and oriented with north up and east to the left. S/N is calculated for an aperture of a resolution element in diameter to the azimuthal variations. \label{Fig2}}
	\end{center}

\end{figure}

Figure \ref{Fig1} shows the disk image of HD 15115 after the KLIP processing. The upward direction indicates the position angle (PA) of 0$^\circ$, i.e. north. Figure \ref{Fig2} shows the signal-to-noise ratio (S/N) map. We defined the signal as the flux in the resolution sized apperture, meanwhile the noise as the azimuthal standard deviation of the signal avoiding the region of the spiders passing. We detected the debris disk from 1.2" to 2.7" on the eastern side and from 1.2" to 3.2" on the western side above the 2$\sigma$ level relative to the sky. The maximum of S/N is 9.0 at 2.0". Because the coronagraph mask is located at 2.5" from the edge of the frame, the outside of the western side of the disk is truncated owing to the limited field of view. We confirm that the disk has a bow-shaped structure (\cite{deb08}) and an asymmetric SB (\cite{kal07}). 


\begin{figure}
\begin{center}
\FigureFile(80mm,80mm){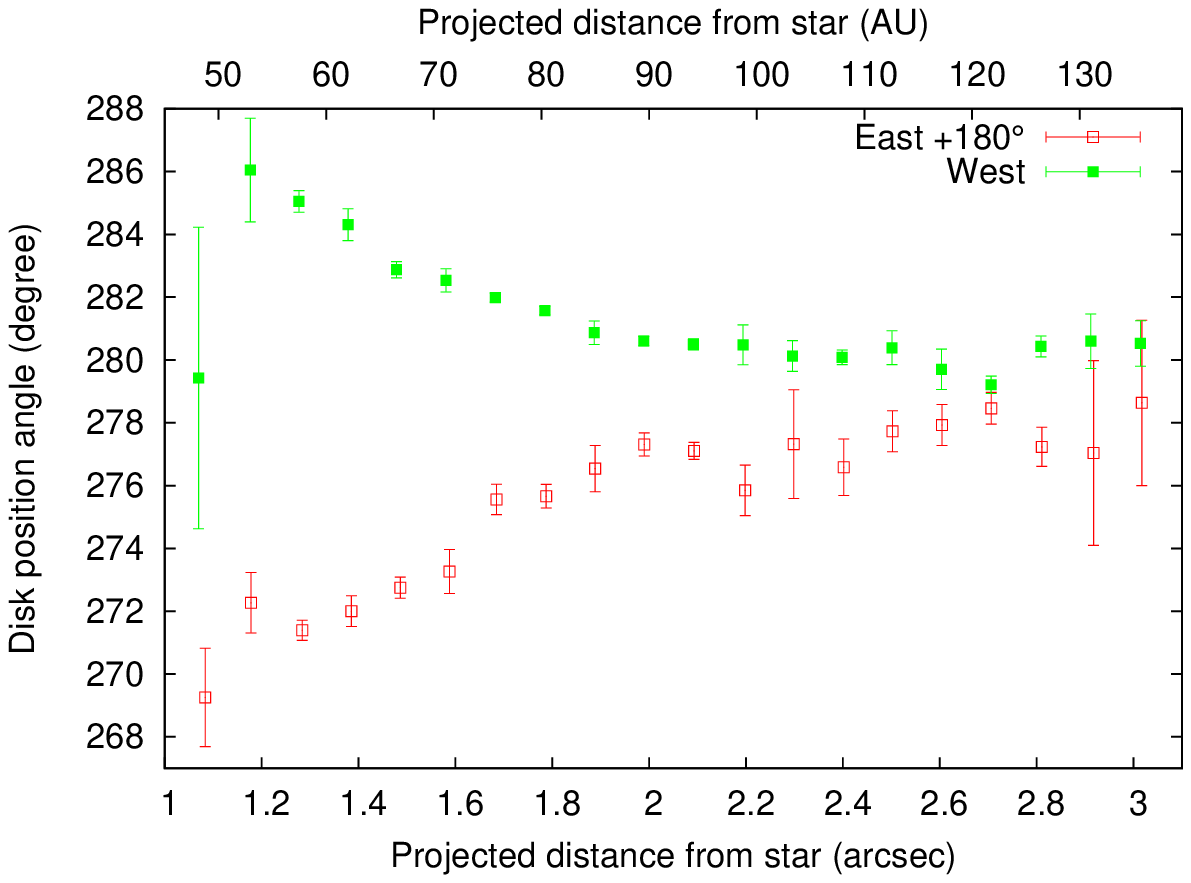}
\caption{Disk position angles of eastern and western sides as a function of projected distance from central star. The slopes of both sides seem to be shallow beyond 2". \label{Fig3}}

\FigureFile(80mm,80mm){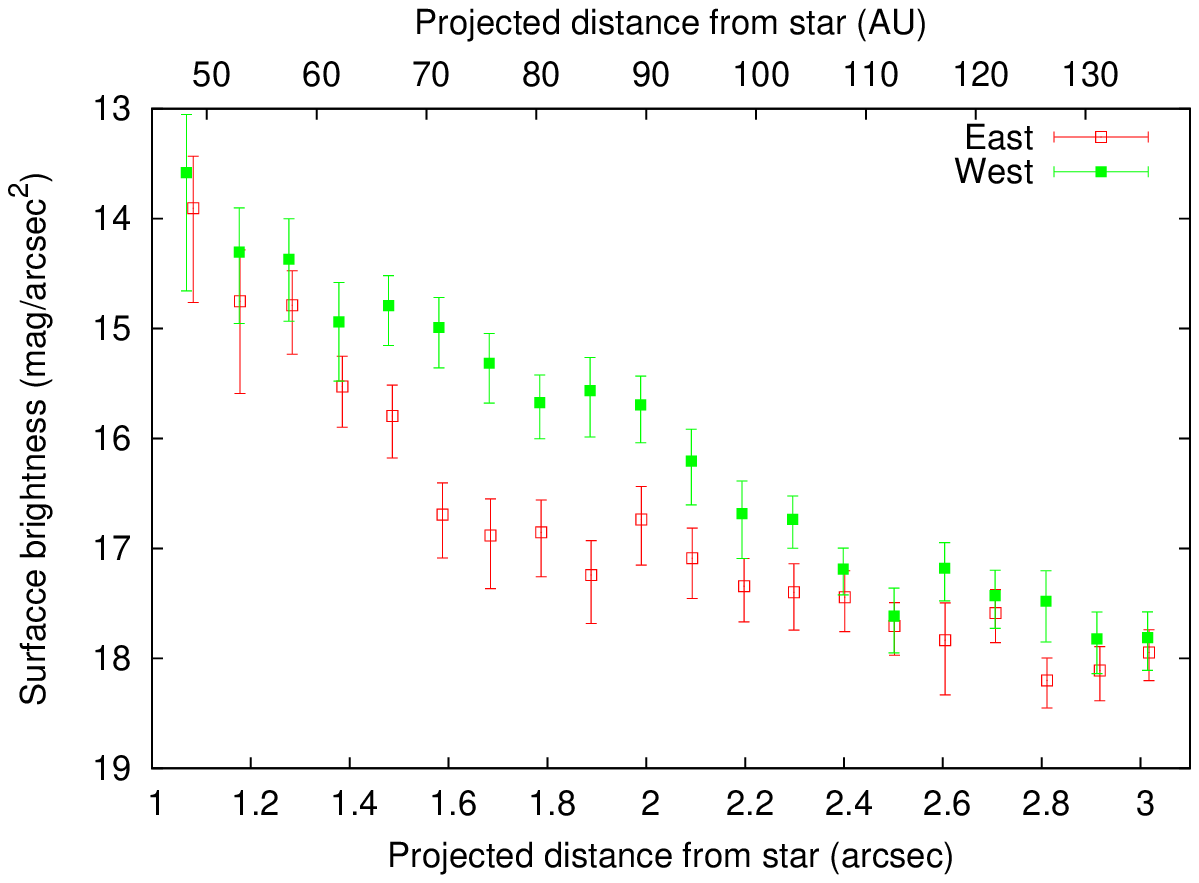}
\caption{Disk surface brightness as a function of projected distance from central star. Self-subtraction correction was applied. The error contains the uncertainty of self-subtraction. The slopes of both sides seem to break at $\sim$2". \label{Fig4}}

\FigureFile(80mm,80mm){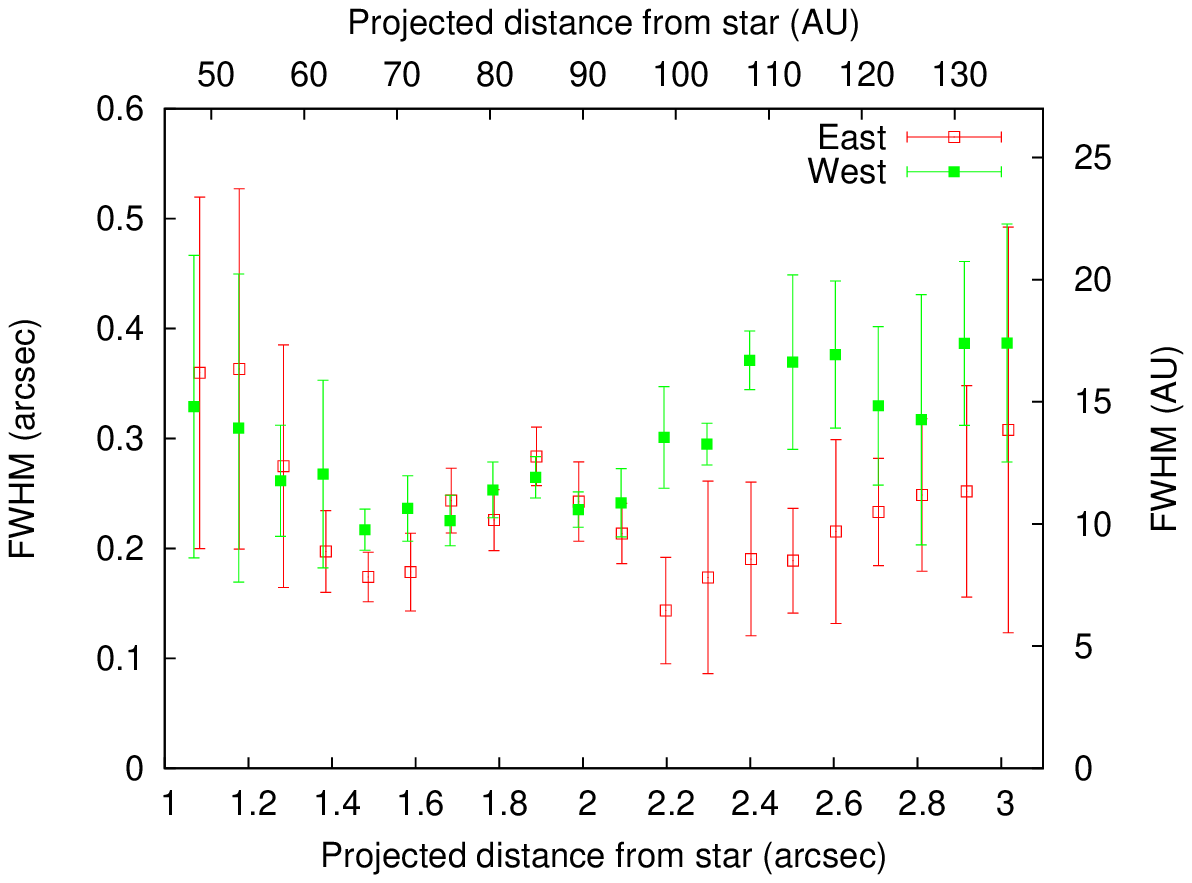}
\caption{Disk FWHM as a function of projected distance from central star. The FWHMs are measured as the thicknesses perpendicular to the orientation of the disk extension. The correction of self-subtraction was applied. Small bumps appear at $\sim$1.9" on the eastern and western sides.\label{Fig5}}
\end{center}
\end{figure}

 For comparison with previous studies, we measured the disk PA, SB, and FWHM as a function of the projected distance. We combined all the science frames with PAs that had a -9.5$^\circ$ offset relative to north (up), so that the disk on the combined image lies almost parallel to the horizontal axis. We fit a Gaussian function vertically along the columns at 1-pixel intervals. The disk PA, SB, and FWHM for each interval were obtained from the peak location, peak value, and dispersion of the Gaussian function, respectively. Then each peak location was transformed into polar coordinates yielding the projected distance and PA. Averaging the PA, SB, and FWHM for every five points (corresponding to the FWHM of the PSF) yielded the evaluation values, while the standard deviations yielded the errors. The errors in the SB also include the uncertainty of self-subtraction. 
Figures \ref{Fig3}--\ref{Fig5} show the measured disk PA, SB, and FWHM, respectively, for the eastern and western sides. The difference in the SB between the two sides is about 0.5--1.5 magnitudes at each projected distance. The flux drop appears at 1.6"--1.9" in the eastern side. The disk SBs are consistent with those of the Keck H band image (\cite{kal07}) with 0.5 magnitude except for the flux drop. It also seems that there are break points of the PA and SB near 2" on both sides. On the other hand, there seems to be a bump at $\sim$2" in the FWHM. Table \ref{Tbl1} shows the SB power-law indices for the inner and outer 2.0"- regions. Our values agree with most of the previous measurements (\cite{kal07}, \cite{deb08}, and \cite{rod12}) within the 1.5 sigma uncertainty.


\begin{table}
\caption{SB Power-law Indices}
\begin{center}
\begin{tabular}{ccccccl}
\hline\hline
Band &Side & Inner Index& & Outer Index& &Reference\\
\hline
H&East &5.1$\pm$0.9 &(1.0-2.0")&2.6$\pm$0.7  &(2.0-3.0") &This Work\\
&West &2.8$\pm$0.7 &(1.0-2.0")&3.4$\pm$0.9  &(2.0-3.0") &\\
\hline
H&East &4.4 &(0.7-2.3")& & &\citet{kal07}\\
&West &3.7 &(0.7-2.3") & & &\\
\hline
J&East &2.14$\pm$0.06&(0.7-3.0")& & &\citet{deb08}\\
&West &1.4$\pm$0.1 &(0.7-1.8") & 3.56$\pm$0.06 &(1.8-3.7")&\\
\hline
Ks&East&5.27$\pm$0.41 &(1.2-1.4")& & &\citet{rod12}\\
& &1.1$\pm$0.07 &(1.4-2.1") & & &\\
&West &1.75$\pm$0.05 &(1.2-1.8")&4.40$\pm$0.24  &(1.8-2.4") &\\
\hline

\end{tabular}
\end{center}
\label{Tbl1}

\end{table}

\section{Disk Geometry}

\subsection{Existence of the Inner Hole in the Disk}

 It appears in Figure \ref{Fig1} that both sides of the disk within 2" of the star curve in the northern direction. This indicates that the disk has an inner hole. If the disk had no inner holes, the projected disk image would be on a straight line even if it is not perfectly edge-on. Not only in the H band but also in the Ks and J bands (\cite{deb08}, \cite{rod12}), the break point of the PA and the SB appears at 2", which supports the existence of an inner hole with 2" radius. This is because we see the rim of the inner hole at $<$2", while we see the inside of the disk at $>$2". The bump of the FWHM at 2" (Figure \ref{Fig5}) also may support the presence of an inner hole, because the front and back side of the disk overlap at the turning point of the disk. The existence of the inner hole at 2" is consistent with \citet{maz14}. \citet{rod12} also noted the existence of the gap, the radius of which is  $R_{\rm in}$=1.1". This value is determined by the drop-offs in the western SB in the Ks and L' bands near 1.1", while we cannot confirm the presence of this signature due to low S/N.


\subsection{Model Fitting}
\begin{figure}
\begin{center}
\FigureFile(80mm,80mm){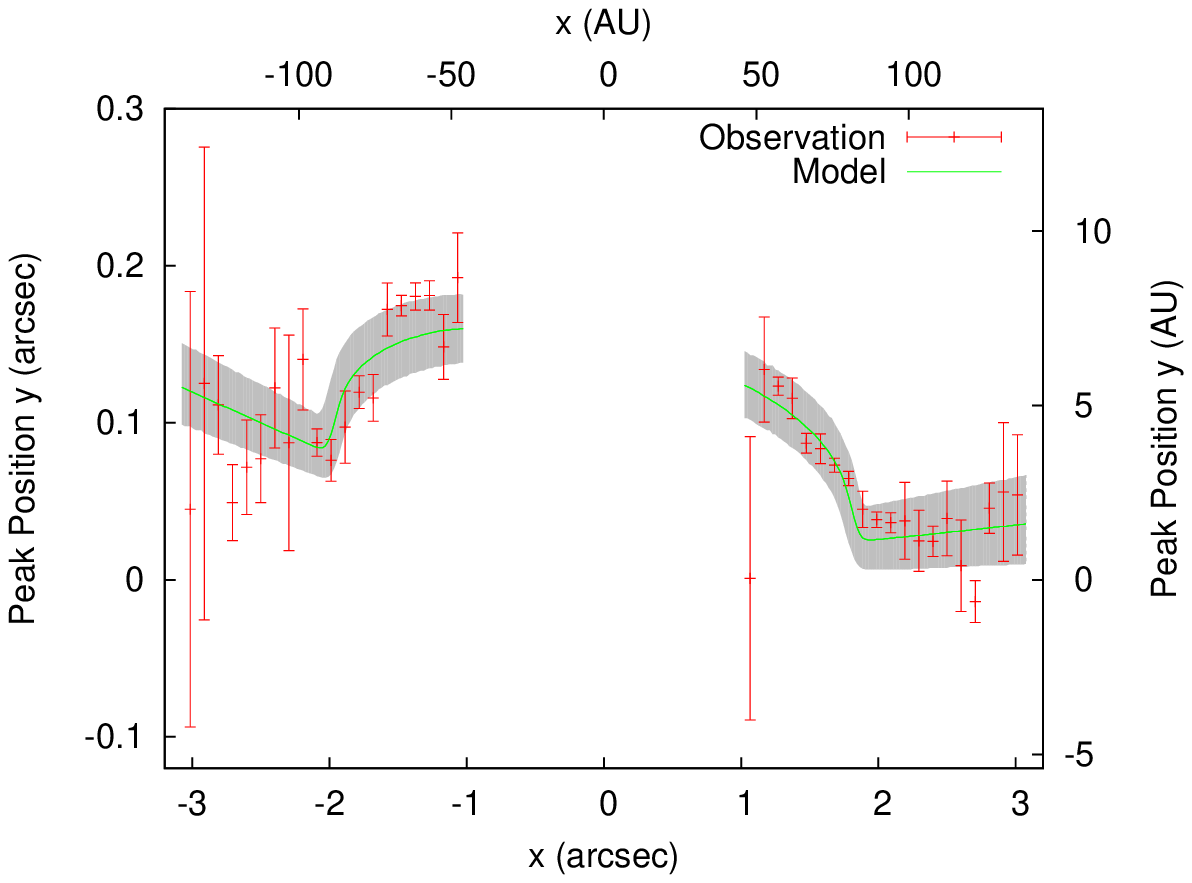}\\
\FigureFile(80mm,80mm){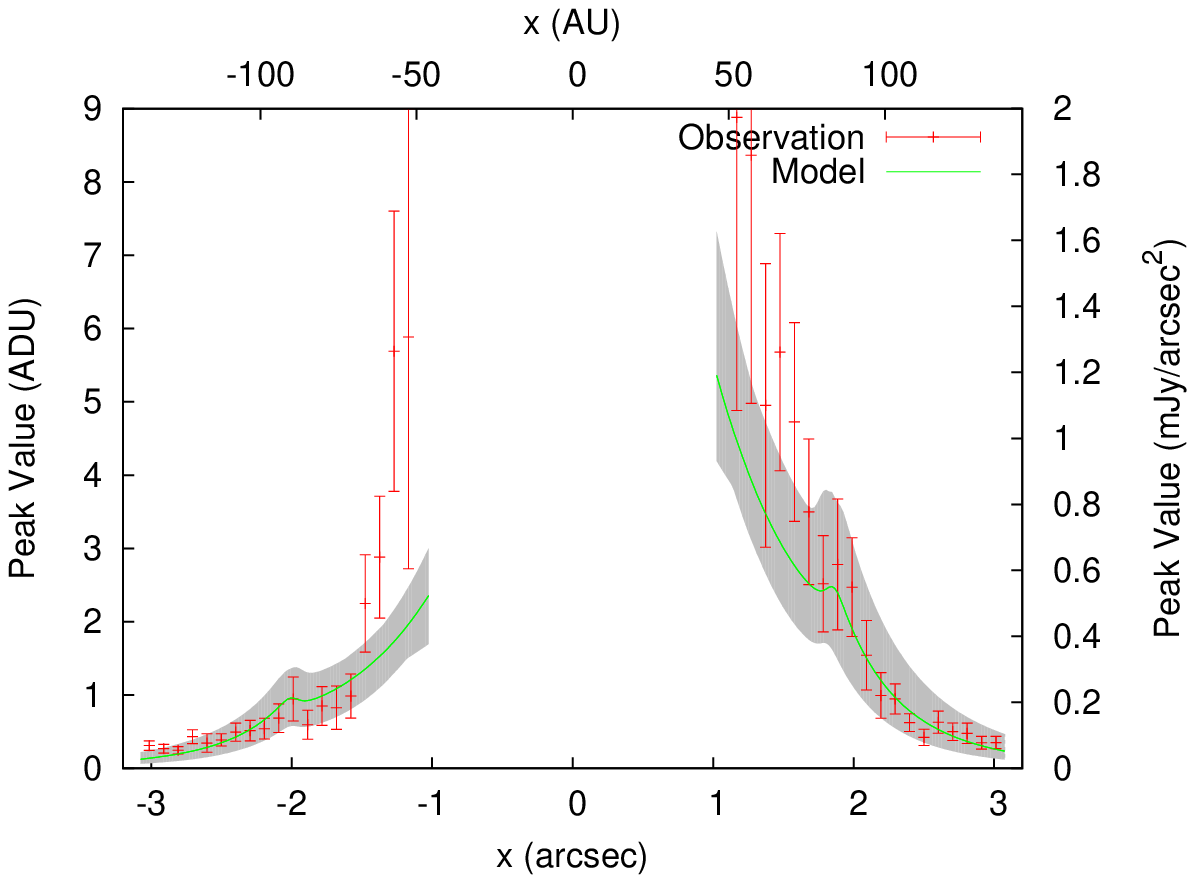}\\
\FigureFile(80mm,80mm){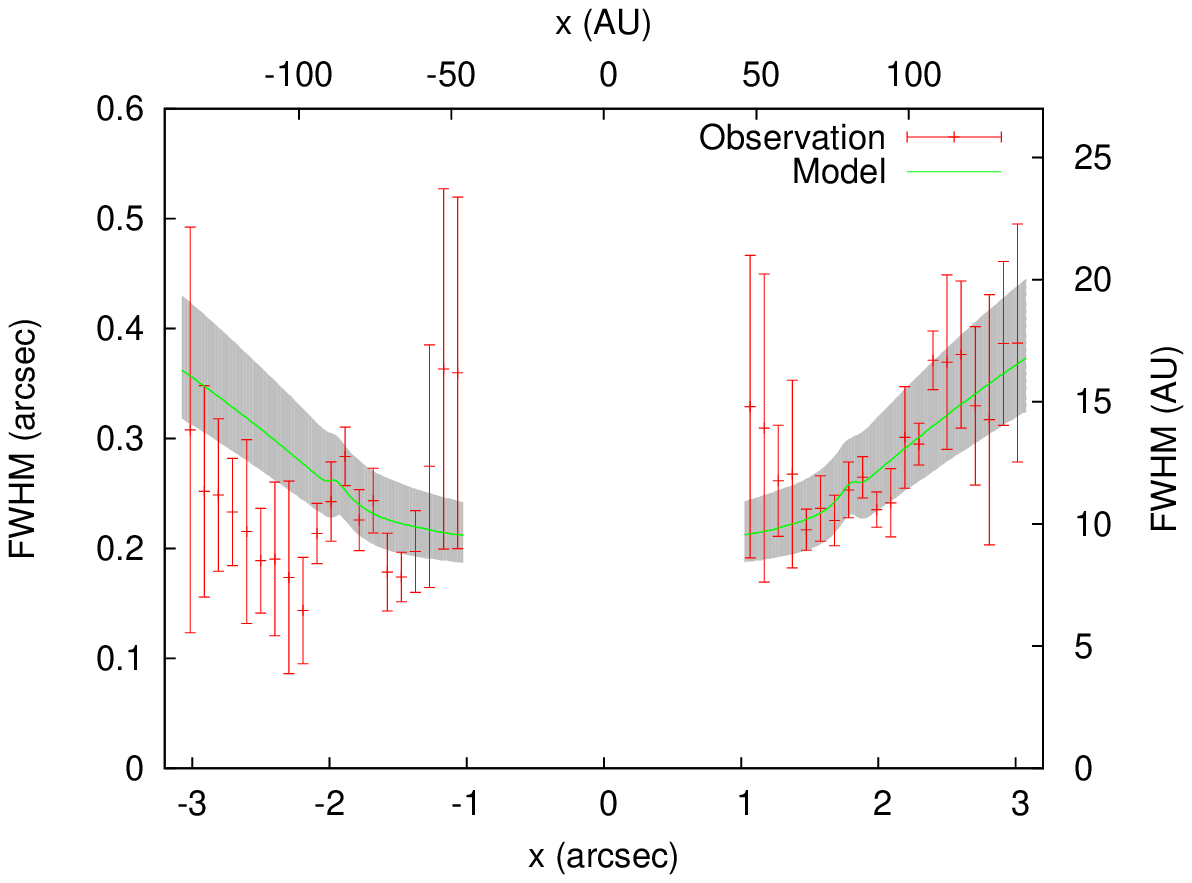}
\end{center}
\caption{Disk peak position (top), peak value (middle) and FWHM (bottom) as a function of the projected horizontal distance $x$. Dots with error bars represent observational results; solid lines represent model profiles. Gray areas represent range in which the parameters fluctuate within errors shown in Table \ref{Tbl2}. \label{Fig6}}
\end{figure}

\begin{figure}
\begin{center}
\FigureFile(100mm,100mm){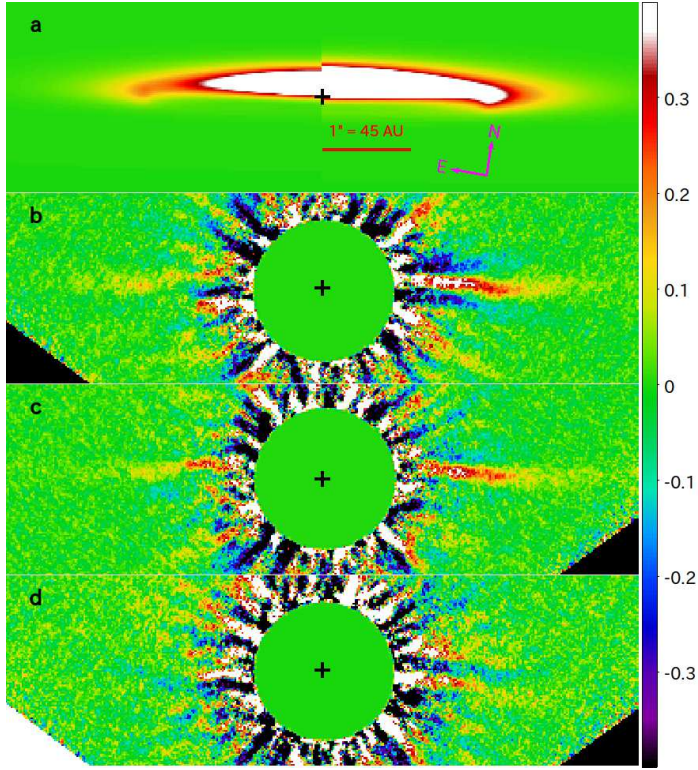}
\end{center}
\caption{Intensity images of model disk and observation. (a) Model disk with best-fit parameters shown in Table \ref{Tbl2}. The image was convolved by the Gaussian function  with FWHM=0.1" corresponding to the PSF size. (b) Recovered model disk image with KLIP. (c) Observational result of HD 15115 real disk. (d) Subtracted image of (c) - (b).  All images have a same brightness scale in units of  ${\rm mJy/arcsec^2}$. \label{Fig7}}
\end{figure}

The bow-shaped structure indicates that the debris disk with an inner hole is highly inclined, and part of the disk is bright from forward scattering of dust (e.g. \cite{cur12}, \cite{rod12}). We use an inclined annulus disk model to probe the disk geometry. We assume that the dust density has a Gaussian distribution in the vertical direction and a power-law profile in the radial direction. The disk thickness increases in proportion to the radius. Thus, the density distribution of the dust at the radius $r$ and height $h$ is described as follows:
\begin{equation}
\left\{ 
\begin{array}{l}

  \rho_{r,h}\propto\left(\frac{r}{R_{\rm in}}\right)^{-q}\exp\left(-\frac{h^2}{{2\left(\sigma_{h}\cdot\frac{r}{R_{\rm in}}\right)^2}}\right) \ \  ( \rm if \ \it r\geq R_{\rm in})\\
 \rho_{r,h}=0 \ \  \ \  \ \  \ \  \ \  \ \  \ \  \ \  \ \  \  \ \  \ \  \ \  \ \  \ \  \ \  \ \ \ \ \ ( \rm if  \ \it r<R_{\rm in})
\end{array}
\right.,
\end{equation}
where $R_{\rm in}$ is the disk inner radius, and $\sigma_{h}$ is the vertical scale height at $r=R_{\rm in}$. Then we inclined the model disk and varied the location of the center of the disk. Assuming that the light scattered by dust follows the Henyey-Greenstein phase function, the surface brightness $F(r,h)$ is described as 
\begin{equation}
F(r,h)=A\cdot\frac{1}{d^2}\cdot\rho_{r,h}\cdot\frac{1-g^2}{(1+g^2-2g\cos\alpha)^{3/2}} ,
\end{equation}
where $A$ is the scaling factor, $d$ is the distance from the star, $g$ is the asymmetry parameter, and $\alpha$ is the scattering angle. To examine the similarity between the observation and the model, we projected the model disk on a two-dimensional plane $(x,y)$ and probed the disk geometry. We derived the peak position, peak value, and FWHM along the vertical direction with the errors at 5-pixel intervals (Figure \ref{Fig6}) and compared the data with the model. We used a minimum $\chi^2$ fitting to search for the best parameters. A 68\% confidence level was calculated by satisfying $\chi^2\leq\chi_{\rm min}^2+1$, where  $\chi_{\rm min}^2$ is the minimum of $\chi^2$. The scaling factor $A$ was determined separately for the eastern and western sides.  The result of the minimum $\chi^2$ fitting is shown in Table \ref{Tbl2}. The best fitted disk has an inner radius $R_{\rm in}$ of 1.91$\pm$0.03" (86.0$\pm$1.2 AU), an inclination $i$ of 86.3$\pm$0.3$^\circ$, a Henyey-Greenstein parameter $g$ of 0.65$\pm$0.10, a power-law index $q$ of the radial dust density of 2.8$\pm$0.5, a vertical scale height $\sigma_{h}$ of 0.058$\pm$0.007" (2.6$\pm$0.3 AU), and a disk PA of 278.63$\pm$0.08$^\circ$. This model fitting also suggests that the disk is off-centered toward the east and that the scaling factor is larger on the western side than on the eastern side. The separation between the star and the disk center, $dr$, is 0.11$\pm$0.04" (5.1$\pm$1.8 AU); thus, the eccentricity of the disk relative to the star, $e=dr/R_{\rm in}$, is 0.06$\pm$0.02. The PA of the disk center relative to the star, $\theta$, is 70$\pm$30$^\circ$. The ratio of $A$ between the eastern and western sides, $A_{\rm west}/A_{\rm east}$, is 2.15$\pm$0.15. Figure \ref{Fig6} shows the disk peak position, peak value, and FWHM obtained using the parameters best fitted by the observation data. We injected the model disk with the best-fit parameters into the science images perpendicularly to the real disk, and then recovered the model disk (Figure \ref{Fig7}). The bottom of Figure \ref{Fig7} shows the image by subtracting the recovered model image from the observational result. The residual disk signal has $<2\sigma$ for S/N, which can be regarded as the level of sky noise. Thus, the model disk is considered a good match with the observation.

\begin{table}

\caption{Best-Fit Model Parameters}
\begin{center}
\begin{tabular}{ll}
\hline\hline
$R_{\rm in} $ &86.0$\pm$1.2 AU\\
$i$ &86.3$\pm$0.3$^\circ$\\
$g $& 0.65$\pm$0.10\\
$q$ & 2.8$\pm$0.5\\
$\sigma_h$ & 2.6$\pm$0.3 AU\\
disk PA & 278.63$\pm$0.08$^\circ$\\
$dr$ & 5.1$\pm$1.8 AU\\
$A_{\rm west}/A_{\rm east}$ &2.15$\pm$0.15\\
\hline

\hline
\end{tabular}
\end{center}
\label{Tbl2}

\end{table}


\section{Discussion}
The model fitting suggests that the disk is off-centered toward the east. This means that the star is closer to the western side of the disk. If the axisymmetric disk purely off-centered by 0.1", the difference of SB at each edge of inner hole is only $\sim$0.2 mag. 
Thus, the dust amount ratio of 2.2 between the western and eastern side is also necessary to explain the asymmetric SB even the western side is closer to the star. We discuss these two characteristics.


\subsection{Disk Inner Hole}
\subsubsection{Radiation pressure}
 The origin of inner hole might be planets. The relevant forces acting on dust grains are radiation pressure and gravity. Hence, the radius of dusts blown out by radiation pressure, $s_{\rm blow}$, can be written as
\begin{equation}
s_{\rm blow}=\frac{3L_*}{8\pi c GM_* \rho}=1.16\left(\frac{\rho}{\rm 1\ g/cm^3}\right)^{-1}\left(\frac{L_*}{L_{\odot}}\right)\left(\frac{M_*}{M_{\odot}}\right)^{-1}\ \micron
\label{eq:s_blow}
\end{equation}
(\cite{chi09}, \cite{rod14a}), where $c$ is the speed of light, $G$ is the gravitational constant and $\rho$ is the internal density of grain. For HD 15115 ($L_*=3.3\ L_{\odot},\ M_*=1.3\ M_{\odot}$), the grains ($\rho=3\ \rm g/cm^3$) whose size are smaller than $s_{\rm blow}=1 \ \micron$ should be blown out by radiation pressure. Meanwhile, the bound dust grains fall toward the star by Poynting-Robertson drag on the timescale of
\begin{equation}
t_{\rm PR}=7 \left(\frac{L_*}{L_{\odot}}\right)^{-1}\left(\frac{s}{1\ \micron}\right)\left(\frac{r}{100\ \rm AU}\right)^2 \rm Myr
\end{equation}
(\cite{cur09}), where $s$ is the grain radius. The remaining micron-sized grains at 100 AU ($\sim$2") should drop into the star in several Myr, which is shorter than the stellar age. Thus, we consider that only radiation pressure without any planets could not open the hole because Poynting-Robertson drag would fill the hole.

\subsubsection{Eccentric disk inner hole and planet possibility}
 The inner hole of the disk is off-centered toward the east by 0.11" (5.1 AU) relative to the star, corresponding to an eccentricity of 0.06.  Our model fitting suggests an offset at the 3 sigma level, while \citet{maz14} indicated that the inner hole was not off-centered. Off-centered debris disks are not uncommon. For example, Fomalhaut and HR 4796 A also have offsets of 15 AU (\cite{kal05}) and $\sim$5 AU (\cite{tha11}), respectively. A phenomena can be considered as a possible cause: an eccentric planet exists inside the disk, shaping the inner gap. A planet's resonant ring would be asymmetric as a result of the resonant trapping of particles into the planet's exterior mean motion resonances (\cite{wya99}). N-body simulations with an eccentric planet indicate that the planet remove the inner debris region, and remaining particles form an elliptical disk apisdally aligned with the planet (\cite{pea14}).

\begin{figure}
\begin{center}
\FigureFile(80mm,80mm){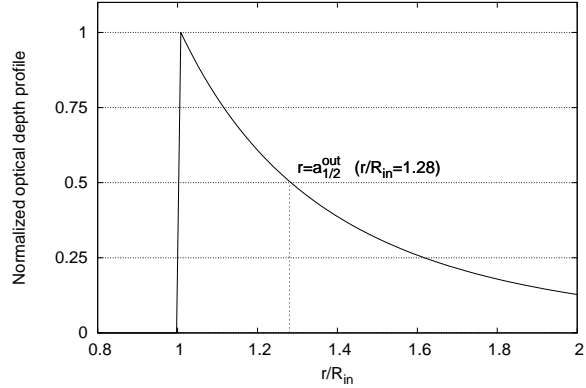}
\end{center}
\caption{Normailzed radial optical depth profile of the best fitted disk model. The profile is scaled to 1 at maximum, as a function of the disk-centered radius normalized by $R_{\rm in}$.   \label{Fig8}}

\end{figure}

\citet{rod14a} presents the estimation of the shepherding planet's maximum mass and minimum semimajor axis from the disk ring width by the N-body simulations. We use their results to estimate the planet's maximum mass and minimum semimajor axis. The planet's maximum mass,  $M_{\rm p, max}$, can be derived by
\begin{equation}
M_{\rm p, max}/M_{\rm Jup}=\left(\frac{\rm nFWHM -(0.107 \pm 0.032)}{0.019 \pm 0.0064}\right)\left(\frac{M_*}{M_{\odot}}\right)
\label{eq5}
 \end{equation}
where $\rm nFWHM$ is FWHM of the profile of face-on optical depth as a function
of radius normalized by the disk semimajor axis, defined as 
\begin{equation} 
{\rm nFWHM}=\frac{a_{1/2}^{\rm out} - a_{1/2}^{\rm in}}{a_{\rm peak}}.
 \end{equation}
Here, $a_{\rm peak}$ is the radius of the optically thickest disk region, and $a_{1/2}^{\rm out}$ and $a_{1/2}^{\rm in}$ are the outer and inner radii where the radial optical depth is half of the peak optical depth. Because we do not measure the inner radial  slope in the model fitting, we use only the outer radial slope to derive $\rm nFWHM$. Figure \ref{Fig8} shows the radial optical depth profile of the best fitted model disk. Since the profile decreases with power of $q$, the radius at the half of maximum, i.e. $a_{1/2}^{\rm out}$ can be derived by   
\begin{equation}
\left(\frac{a_{1/2}^{\rm out}}{R_{\rm in}}\right)^{-q}=\frac{1}{2}.
\end{equation}
Assuming $a_{1/2}^{\rm in}=a_{\rm peak}=R_{\rm in}$, we obtain
 \begin{equation}
{\rm nFWHM}=\frac{a_{1/2}^{\rm out}-R_{\rm in}}{R_{\rm in}}=\frac{2^{1/q}R_{\rm in}-R_{\rm in}}{R_{\rm in}}=2^{1/q}-1.
\label{eq8}
 \end{equation}
The equations (\ref{eq5}) and (\ref{eq8}) with $q=2.8\pm0.5$ (Table \ref{Tbl2}) yield $\rm nFWHM=0.28\pm0.04$ and  $M_{\rm p, max}=12\pm5\ M_{\rm Jup}$. Meanwhile, the minimum semimajor axis, $a_{\rm p, min}$, can be derived by
\begin{equation}
 a_{\rm p, min}=\frac{R_{\rm in}}{1+10.23(M_{\rm p, max}/M_*)^{0.51}}.
\end{equation} 
Substituting $R_{\rm in}=86.0\pm1.2\ \rm AU$ and $M_*=1.3\ M_{\odot}$ gives $a_{\rm p, min}=45\pm 5\ \rm AU\ (1.0\pm0.1")$. \citet{rod12} reported non-detection of point sources in the L' band. The 5$\sigma$ detection limit ranged from magnitude $\sim$15 to $\sim$18 with separations of 0.6" and 2.4", respectively. Using their detection limit with the COND model (\cite{bar03}), we calculated the detection probabilities in their L' band observation.
Given an nFWHM of 0.28, the maximum planet mass is $12\ M_{\rm Jup}$ at a minimum semimarjor axis of 45 AU.
Such a planet would have been detected by Rodigas et al. 2012, assuming the planet was at a favorable projected separation during their observations and that the star has an age of a few 10 Myr.
Additional deeper observations would help constrain the physical properties of the unseen planet.

\subsection{Difference in the amount of dust on either sides}

The model fitting suggests that the brightness on the western side is 2.2 times larger than that on the eastern side. When we consider an optically thin disk with the same cross-sectional of dust profile, this value corresponds to the ratio of the dust number between the eastern and western sides. To account for the difference in the amount of the dust, three phenomena are cited as a possible cause: 1) A dramatic planetesimal impact has recently occurred on the western side, 2) A Jupiter-mass planet captures grains at its Lagrangian points, or 3) ISM gas blows out grains from the eastern side to the west.

\subsubsection{Dramatic planetesimal impact}
\citet{hah10} proposed that one possible explanation for a disk's asymmetry is a recent planetesimal collision. \citet{sta14} claimed to see evidence for such an impact in the HD 181327 debris disk system. The total dust mass of HD 15115 is $\sim0.05\ M_\earth$, which was calculated from the 850 \micron\ flux (\cite{wil06}). Assuming that the number of dust particles of any size on the western side is 2.2 times greater than the number on the eastern side, we obtain $M\sim0.02 M_\earth$ for the excess mass. Here, we consider the scenario of dust production by the collision of two planetesimals (mass: $M_1$, $M_2$; radius: $R_1$, $R_2$). The two bodies collide with each other and then a combined body remains (mass: $M_{\rm comb}$; radius: $R_{\rm comb}$) with emitting dust grains (mass: $m_{\rm d}=M_1+M_2-M_{\rm comb}$). We can define four equations:
\begin{equation}
U=\frac{3GM_1^2}{5R_1}+\frac{3GM_2^2}{5R_2}-\frac{3GM_{\rm comb}^2}{5R_{\rm comb}},
\label{eq10}
\end{equation}
\begin{equation}
E_{\rm col}=\frac{1}{2}\frac{M_1M_2}{M_1+M_2}v_{\rm imp}^2,
\label{eq11}
\end{equation}
\begin{equation}
v_{\rm imp}^2=v_{\rm disp}^2+\frac{2G(M_1+M_2)}{R_1+R_2},
\label{eq12}
\end{equation}
\begin{equation}
v_{\rm disp} \sim \sqrt{6} h v_{\rm Kep},
\end{equation}
where $v_{\rm imp}$ is the impact velocity, $v_{\rm disp}$ is the velocity dispersion, $h$ is the disk aspect ratio, and $v_{\rm Kep}$ is the orbital velocity. The first equation is the required energy to produce dust grains, while the second equation is the collisional energy. The third equation indicates that the impact velocity is increased by the energy conversation. The fourth equation means that the dispersion velocity can be derived with the disk aspect ratio and the orbital velocity (\cite{qui07}). When we use $h=\sigma_h/R_{\rm in}=0.030$ and $v_{\rm Kep}=3.7$ km/s at $r=86$ AU, we obtain $v_{\rm disp}=0.27$ km/s. For the collision of several 1000 km-size bodies with internal density of $\rho=3$ g/$\rm cm^3$ (typical value for 1000 km-size bodies in the solar system, e.g. Europa, Io), equation (\ref{eq12}) indicates that $v_{\rm imp}$ grows to several km/s. Using the equations (\ref{eq10}),  (\ref{eq11}), (\ref{eq12}), and $U=E_{\rm col}$ with these values of $v_{\rm disp}$ and $\rho$, we found that the collisional two bodies would have $M_1=M_2\sim0.02\ M_\earth$ and $R_1=R_2\sim2000$ km to produce dust grains of $m_{\rm d}=0.02 M_\earth$.  


Then, we consider the collisional timescale of two giant planetesimals with the theoretical approach presented by \citet{wya99}. The collisional timescale of particles with a diameter of $D$ at a radius of $r$ from the star can be written as
\begin{equation}
t_{\rm coll}(D,\ r)\approx t_{\rm coll}(D_{\rm typ},\ r)\left(\frac{D_{\rm cc}(D)}{D_{\rm typ}}\right)^{3q_{\rm s}-5},
\end{equation}
where $D_{\rm typ}$ is the diameter of the dominant particles in the disk, $D_{\rm cc}(D)$ is the minimum diameter to have a catastrophic collision for the planetesimal with $D$ in diameter, and $q_{\rm s}$ is a power index of the diameter distribution. 
When we use $q_{\rm s}=11/6$, which is derived from collisional cascade (\cite{doh69}) and assume $D_{\rm cc}(4000\ \rm km)=4000\ \rm km$, the collisional timescale is 
\begin{equation}
t_{\rm coll}(4000\ {\rm km},\ r)\sim2\times 10^6 \left(\frac{D_{\rm typ}}{1\ \micron}\right)^{-1/2}t_{\rm coll}(D_{\rm typ},\ r).
\end{equation}
Here, the collisional timescale of the typical particle diameter can be described by
\begin{equation}
t_{\rm coll}(D_{\rm typ},\ r)=\frac{t_{\rm per}(r)}{4 \pi \tau (r)},
\end{equation}
where $t_{\rm per}(r)$ is the average orbital period of particles at $r$ and $\tau (r)$ is the disk's face-on optical depth. Assuming that the disk consists of $D_{\rm typ}$ diameter particles only, we obtained  $t_{\rm per}(r)$ and $\tau (r)$ at $r=R_{\rm in}$, as
\begin{equation}
t_{\rm coll}(4000\ {\rm km},\ 86\ {\rm AU})\sim 240\ {\rm Myr} \left(\frac{D_{\rm typ}}{1\ \micron}\right).
\end{equation}
 Considering $D_{\rm typ}$ is $2\ \micron$, which corresponds to the blow out diameter (Equation (\ref{eq:s_blow})), $t_{\rm coll}(4000\ {\rm km},\ 86\ {\rm AU})$ should be on the order of 100 Myr. Therefore, it is possible that a dramatic planetesimal impact has produced the extra amount of dust if the age is several 100 Myr.   




\subsubsection{Grain trapping at Lagrangian points by a Jupiter-mass planet}
 When a planet exists in the system, it may capture dust particles at its Lagrangian points. Among those points, L4 and L5 are stable enough to retain many grains (\cite{lio99}; \cite{ert12}; \cite{the12}). If the L4 or L5 point comes to the western side, then an excess amount of dust should be seen. In that case, the planet is also located on the western side. The stability of the L4 or L5 point depends on the eccentricity of the planet and on the masses of the star, the planet, and the Trojan body (\cite{mar90}; \cite{dvo07}). \citet{mar90} defined the mass parameter as
\begin{equation}
P=\frac{(m_2+m_3)}{M}+\frac{m_2\cdot m_3}{m_1^2}+O\left(m_2^3 \cdot \frac{m_3}{m_1^4}\right),
\end{equation}
where $m_1$, $m_2$, $m_3$, and $M$ are the the masses the star, the planet, the Trojan particles, and the total mass of the system, respectively. Assuming that the planet has an eccentricity of $e=0.06$, which is equal to the eccentricity of the disk inner hole, $P$ must be smaller than 0.025 to keep the system stable. On this condition, the mass of the planet has only to be smaller than 35 $M_{\rm Jup}$.  In addition, the Ks - L' disk color suggests that the western side of the disk is composed of smaller grains (1--3 $\micron$) of dust than the eastern side (\cite{rod12}). Small grains are subjected to move radially through the system by Poynting-Robertson drag and therefore can be easily trapped into resonance by the planet (\cite{ert12}). If a planet exists in the HD 15115 system, many small grains might be gathering around the L4 and L5 points. In conclusion, this scenario could occur, however, simulations customized to HD 15115 system are necessary to confirm quantitatively.


\subsubsection{ISM interaction}
 \citet{deb09} first suggested that the local ISM could cause the brighter disk in the western side than in the eastern side. The space motion of HD 15115 to the south east (PA = 120$^\circ$, 100 mas/yr) nearly parallel to its disk major axis (PA = 99$^\circ$) is consistent with this suggestion. Their model indicates that the relative velocity between the ISM gas and HD 15115 would be $\sim$30 km/s to cause the $\sim$1 mag brightness difference at J band.  \citet{rod12} showed the redder Ks -L' disk color in the eastern side than western side to support this ISM interaction. Small grains (1--3 $\micron$) would be blown back to the west by the ISM wind, while large particles ($>3 \micron$) would remain in their initial orbits. Our H band disk is 0.5-1.5 mag brighter in the western side than in the eastern side. This asymmetry is nearly same with Ks or J band. The V band image shows the stronger asymmetry; the western side of the disk extends to $>$ 550 AU, whereas the eastern side of the disk extends to $\sim$315 AU (\cite{kal07}). Such asymmetries observed are consistent with the idea of the ISM interaction.

Meanwhile, our model fitting with the H band image suggests that the inner hole does not shift to the west but the east. It is considered that the inner part of the disk has large grains (\cite{str06}) which are not affected by interactions with the ISM. Therefore, the ISM would hardly affect the behavior of the inner hole. Considering the extremely asymmetric V band disk at $>$ 300 AU, ISM interaction would only blows small grains in the outside, while the grains in the inner region ($\lesssim$ 100 AU) are large enough to be unperturbed. 




\section{Summary}
We obtained an H-band image of the disk around HD 15115 using Subaru/IRCS. The main results are summarized as follows:
\begin{enumerate}
 \item We detected the nearly edge-on disk around HD 15115 in the H band.
 \item A bow-shaped structure appears interior to 2" (90 AU) from the star. The bow-shaped structure indicates that the debris disk has an inner hole and is highly inclined, and part of the disk is bright because of forward scattering of dust. It is deduced that the radius of the inner hole is 1.9" (86 AU). The disk inner gap is off-centered toward the east by 0.1" (5 AU) relative to the star. HD 15115 might have an eccentric Jupiter-mass planet that is sculpting the inner edge of the disk.
 \item The disk SB is $\sim$0.5--1.5 mag brighter on the western side than on the eastern side. We consider that the western side has 2.2 times more dust than the eastern side. The difference in the amount of dust may be due to the accumulation of grains at the Lagrangian points of the planet. If the stellar age is over $\sim$200 Myr, a massive planetesimal impact possibly causes the asymmetric dust amount. ISM interaction is also considered as a possible explanation, nevertheless it cannot affect larger grains in the vicinity of the inner hole.

\end{enumerate}
\bigskip

This work is based on observations made with the Subaru Telescope, which is operated by the National Astronomical Observatory of Japan. We thank telescope operators and AO188 team for performing observations. We thank Thayne Currie, the PI of this observation, for providing us with data and the LOCI reduction pipeline and having deep discussion. We also thank Mihoko Konishi, Yusuke Ito, Kodai Yamamoto, Yoshihiro Kuwada and other laboratory members in Osaka University for meaningful discussion.




\bigskip



\end{document}